\documentstyle[prl,aps,twocolumn]{revtex}

\catcode`\@=11

\def\maketitle2{\par 
\begingroup
\let\cite\@bylinecite
\def\thefootnote{\fnsymbol{footnote}}%
\twocolumn[\@maketitle2\vskip2pc]%
\thispagestyle{plain}\@thanks
\endgroup
\def\thefootnote{\arabic{footnote}}%
\setcounter{footnote}{0}%
\let\maketitle2\relax \let\@maketitle2\relax
\let\@thanks\relax \let\@authoraddress\relax \let\@title\relax
\let\@date\relax \let\thanks\relax \let\@abstract\relax 
\let\@pacs\relax}

\def\abstract#1{\gdef\@abstract{{\par 
\bgroup
\ifdim\prevdepth=-1000pt \prevdepth0pt\fi
\hsize\columnwidth
\dimen0=-\prevdepth \advance\dimen0 by17.5pt \nointerlineskip
\small\vrule width 0pt height\dimen0 \relax}{~~}#1\egroup}}

\def\pacs#1{\gdef\@pacs{{\par 
\bgroup
\hsize\columnwidth \parindent0pt
\ifdim\prevdepth=-1000pt \prevdepth0pt\fi
\dimen0=-\prevdepth \advance\dimen0 by20pt\nointerlineskip
\egroup} PACS numbers:~#1}}

\def\@maketitle2{
\@preprint
\@title
\ifdim\prevdepth=-1000pt \prevdepth0pt\fi
\@authoraddress
\@date
\begin{list}{}{\leftmargin=0.10753\textwidth \rightmargin=\leftmargin
\itemsep=1pc\partopsep=-1pc}
\item\@abstract
\item\@pacs
\end{list}
}

\catcode`\@=12

\begin{document}
\draft
\title{Lyapunov Exponents without Rescaling and Reorthogonalization}	 
\author{Govindan Rangarajan,$^{1,*}$ Salman Habib,$^{2,\dagger}$ and 
Robert D. Ryne$^{3,\#}$} 
\preprint{LA-UR-97-2233}
\address{$^1$Department of Mathematics and Center for Theoretical
Studies, Indian Institute of Science, Bangalore 560 012, India}
\address{$^2$T-8, Theoretical Division, MS B285, Los Alamos National
Laboratory, Los Alamos, NM 87545, USA} 
\address{$^3$LANSCE-1, LANSCE Division, MS H817, Los Alamos National
Laboratory, Los Alamos, NM 87545, USA} 
\date{\today} 

\abstract 
{We present a new method for the computation of Lyapunov exponents
utilizing representations of orthogonal matrices applied to
decompositions of $M$ or $M\tilde{M}$ where $M$ is the tangent
map. This method uses a minimal set of variables, does not require
renormalization or reorthogonalization, can be used to efficiently
compute partial Lyapunov spectra, and does not break down when the
Lyapunov spectrum is degenerate.}

\maketitle2 
\narrowtext 

Chaotic dynamics has been investigated in a very large class of
systems, including astrophysical, biological, and chemical systems,
mechanical devices, models of the weather, lasers, plasmas, and
fluids, to mention a few. Lyapunov exponents provide the single most
important quantitative characterization of the exponential divergence
of initially nearby trajectories, which is the hallmark of
chaos. Recent applications of these exponents include the connection
between chaotic dynamics and transport theory in statistical mechanics
\cite{trans1,trans2} and galactic dynamics \cite{galdyn}.

Several methods exist for computing Lyapunov exponents
\cite{ER,WEA,GPL,HR,HRFC}. However, no single method appears to be
optimal. For example, QR and SVD (singular value decomposition)
methods \cite{WEA,GPL} require frequent renormalization (to combat  
exponential growth of the separation vector between the fiducial and
nearby trajectories) and reorthogonalization (to overcome the
exponential collapse of initially orthogonal separation vectors onto
the direction of maximal growth). The existing continuous versions of
the QR and SVD methods also suffer from the additional disadvantage of
being unable to compute the partial Lyapunov spectrum using a fewer
number of equations/operations than required for the computation of
the full spectrum \cite{GPL}. Further, the continuous SVD method
breaks down when computing degenerate Lyapunov spectra \cite{GPL}. The
symplectic method \cite{HR} is applicable only to Hamiltonian systems
(and a few generalizations thereof) and has proven difficult to extend
to systems of moderate size, though this is possible in principle
\cite{HR2}. It also does not permit easy evaluation of partial
Lyapunov spectra.

The widespread perception that some form of explicit rescaling and
reorthogonalization is necessary lies at the heart of most methods for
computing Lyapunov exponents. In this Letter, we propose a general
method which analytically obviates the need for rescaling and
reorthogonalization. Our new method also does away with the other
shortcomings listed above: A partial Lyapunov spectrum can be computed
using a fewer number of equations as compared to the computation of
the full spectrum, there is no difficulty in evaluating degenerate
Lyapunov spectra, the equations are straightforward to generalize to
higher dimensions, and the method uses the minimal set of dynamical
variables. Since our method is based on exact differential equations
for the Lyapunov exponents, global invariances of the Lyapunov
spectrum can be preserved. 

The key feature of our approach is the use of explicit group
theoretical representations of orthogonal matrices. This results in a
set of coupled ordinary differential equations for the Lyapunov
exponents along with the various angles parametrising the orthogonal
matrices. The system of differential equations is treated as an
initial value problem and solved numerically to obtain the Lyapunov
exponents. In the preferred variant of our method, the equations are
only partially coupled leading to easy evaluation of the incomplete
Lyapunov spectrum. An interesting consequence of our methodology is
the natural separation between ``slow'' (the exponents) and ``fast''
(the angles) pieces in the evolution equations. (This fact can be used
to provide speed-up in numerical implementations.) Since the structure
of the coupled differential equations is of a special form, they may
also turn out to be useful for analytic studies of evolution in
tangent space.

To begin, we consider an $n$ dimensional continuous-time dynamical
system,  
\begin{equation}
{d{\bf z}\over dt} = {\bf F}({\bf z},t)~,
\label{dynsys}
\end{equation}
where ${\bf z}=(z_1,z_2, \cdots ,z_n)$ and ${\bf F}$ is a
$n$-dimensional vector field. Let ${\bf Z}(t)={\bf z}(t)-{\bf z}_0(t)$
denote deviations from the fiducial trajectory ${\bf
z}_0(t)$. Linearizing Eq. (\ref{dynsys}) around this trajectory, we
obtain 
\begin{equation}
{d{\bf Z}\over dt} = {\bf DF}({\bf z}_0(t),t) \cdot {\bf Z}~,
\label{linear}
\end{equation}
where ${\bf DF}$ denotes the $n \times n$ Jacobian matrix.

Integrating the linearized equations along the fiducial trajectory
yields the tangent map $M({\bf z}_0(t),t)$ which takes the initial
variables ${\bf Z}^{in}$ into the time-evolved variables ${\bf
Z}(t)=M(t) {\bf Z}^{in}$ (the dependence of $M$ on the fiducial
trajectory ${\bf z}_0(t)$ is understood). Let $\Lambda$ be an $n
\times n$ matrix given by $\Lambda = \lim_{t \to \infty} (M
\tilde{M})^{1/2t}$, where $\tilde{M}$ denotes the matrix transpose of
$M$. The Lyapunov exponents then equal the logarithm of the
eigenvalues of $\Lambda$ \cite{ER}.

It is clear that $M$ is of central importance in the evaluation of
Lyapunov exponents. Its evolution equation can be easily derived:
\begin{equation}
\frac{dM}{dt} = {\bf DF}~M~.
\label{Mdot}
\end{equation}
Instead of a brute force attack, our purpose is now to write $M$ (or
some combination thereof) in such a way that the resulting evolution
equations are intrinsically well-behaved. One way to do this is to
follow the approach of Ref. \cite{HR} and introduce the matrix
$A \equiv M \tilde{M}$. The evolution equation for $A$ follows from 
(\ref{Mdot}):
\begin{equation}
{dA\over dt} = {\bf DF}~A + A~\tilde{{\bf DF}}~.
\label{Adot}
\end{equation}

The matrix $A$ is symmetric and positive-definite \cite{ER}. Hence it
can be written as an exponential of a symmetric matrix $B$ \cite{GL}:
$A = \hbox{e}^B$. Furthermore, any symmetric matrix can be
diagonalized using orthogonal matrices \cite{GL}. Thus, $A =
\hbox{e}^{O D O^{-1}}$, where $O$ is an $n \times n$ orthogonal
matrix, $D$ is an $n \times n$ diagonal matrix and $O^{-1} =
\tilde{O}$. From standard properties of matrix exponentials, it
follows that $A = O \hbox{e}^D O^{-1}$. There is no need for
rescaling since the diagonal matrix $D$ is already in the exponent
(the diagonal elements are just the Lyapunov exponents multiplied by
time). 

To proceed further, we use an easy to obtain explicit representation
of the orthogonal matrix $O$ from group representation theory
\cite{gelfand}. One advantage is that a minimum number of variables is
used to characterize the system: $n(n-1)/2$ in $O$ and a further $n$
variables in $D$, for a total of $n(n+1)/2$. Another advantage is that
numerical errors can never lead to loss of orthogonality. Finally, the
dynamical equations (\ref{Adot}) are solved numerically.

We describe in detail below a variant of this idea which has certain
further advantages. As is well-known \cite{GL}, the matrix $M$ can be
written as the product $M = QR$ of an orthogonal $n \times n$ matrix
$Q$ and an upper-triangular $n \times n$ matrix $R$ with positive
diagonal entries. Substituting this into Eq. (\ref{Mdot}), we obtain:
\begin{equation}
\dot{Q} R + Q \dot{R} = {\bf DF}~QR~,
\end{equation}
where the overdot denotes a time derivative. Multiplying the above
equation by $\tilde{Q}$ from the left and $R^{-1}$ from the right, we
get  
\begin{equation}
\tilde{Q} {\dot Q} + {\dot R} R^{-1} = \tilde{Q}\ {\bf DF}\ Q~.
\label{QReqn}\end{equation}
Note that $\tilde{Q} \dot{Q}$ is a skew(anti)-symmetric matrix for any
orthogonal matrix $Q$ and $\dot{R}R^{-1}$ is still an upper-triangular
matrix.

As before, we now employ an explicit representation of the orthogonal
matrix $Q$ representing it as a product of $n(n-1)/2$ orthogonal 
matrices, each of which corresponds to a simple rotation in the 
$i-j$th plane ($i < j$).  Denoting the matrix corresponding to this
rotation by $O^{(ij)}$, its matrix elements are given by:
\begin{eqnarray}
O^{(ij)}_{kl} & = & 1\ {\rm if} \  k=l \neq i,j~; \nonumber\\
              & = & \cos \phi \ {\rm if} \ k=l=i\ {\rm or} \ j~; 
\nonumber\\
              & = & \sin \phi \ {\rm if} \ k=i,\ l=j~; \nonumber\\
              & = & -\sin \phi \ {\rm if} \ k=j,\ l=i~; \nonumber\\
              & = & 0 \ {\rm otherwise}.
\end{eqnarray}
Here $\phi$ denotes an angle variable. Thus, the $n \times n$ matrix
$Q$ is represented by:
\begin{equation}
Q = O^{(12)} O^{(13)} \cdots O^{(1n)} O^{(23)} \cdots O^{(n-1,n)}~.
\end{equation}
Hence $Q$ is parametrized by $n(n-1)/2$ angles which we denote by
$\theta_i$ ($i=1,\cdots ,n(n-1)/2$). These angles will be
collectively denoted by ${\theta}$.

Since the upper-triangular matrix $R$ has positive diagonal entries,
it can be represented as follows:
\begin{equation}
R = \left( \begin{array}{ccccc} 
e^{\lambda_1} & r_{12} & \cdots &  \cdots & r_{1n} \\
0 & e^{\lambda_2} & r_{23} & \cdots & r_{2n} \\
\vdots & \vdots & \vdots & \vdots & \vdots \\
0 & 0 & 0 & 0 & e^{\lambda_n} 
\end{array}
\right).
\end{equation}
The quantities $\lambda_i$ will be shown to be intimately related to
the Lyapunov exponents. Our final equations will be in terms of the
$\lambda_i$ which already appear in the exponent, thus removing the
need for rescaling. The quantities $r_{ij}$ represent the
supra-diagonal terms in $R$.

Using the above representations of $Q$ and $R$, we obtain:
\begin{equation}
\tilde{Q} \dot{Q} = \left( 
\begin{array}{cccc}
0 & -f_1({\dot{\theta}}) &  \cdots & 
-f_{n-1}({\dot{\theta}}) \\
f_1({\dot{\theta}}) & 0 & \cdots & 
-f_{2n-3}({\dot{\theta}})  \\
\vdots & \vdots & \vdots & \vdots \\
f_{n-1}({\dot{\theta}}) & \cdots 
& f_{n(n-1)/2}({\dot{\theta}}) & 0 
\end{array}
\right)
\end{equation}
and
\begin{equation}
\dot{R} R^{-1} = \left( 
\begin{array}{ccccc}
\dot{\lambda}_1 & r'_{12} & \cdots &  \cdots & r'_{1n} \\
0 & \dot{\lambda}_2 & r'_{23} & \cdots & r'_{2n} \\
\vdots & \vdots & \vdots & \vdots & \vdots \\
0 & 0 & 0 & 0 & \dot{\lambda}_n 
\end{array}
\right).
\end{equation}
Here, each of the $n(n-1)/2$ functions $f_i$ depend (in principle)
on the time derivatives $\dot{\theta}_i$ of all the angles used to
represent $Q$. In fact, they actually depend only on a subset of the
angles. The quantities $r'_{ij}$ are of no concern since they are not
present in the final equations. 

Substituting the above two expressions in Eq. (\ref{QReqn}) we obtain
\begin{equation}
\left( 
\begin{array}{ccccc}
\dot{\lambda}_1 & r''_{12} & \cdots &  \cdots & r''_{1n} \\
f_1({\dot{\theta}})  & \dot{\lambda}_2 & r''_{23} & \cdots &
r''_{2n} \\ 
\vdots & \vdots & \vdots & \vdots & \vdots \\
f_{n-1}({\dot{\theta}})  & \cdots & \cdots & f_{n(n-1)/2}({\dot{\theta}}) 
& \dot{\lambda}_n 
\end{array}
\right) = \tilde{Q}\ {\bf DF}\ Q~.
\label{finaleqn}
\end{equation}
Denoting the matrix $\tilde{Q}\ {\bf DF}\ Q$ by $S$ and comparing
diagonal elements on both sides of (\ref{finaleqn}) one gets:
\begin{equation}
\dot{\lambda}_i = S_{ii}, \ \ \ i=1,2, \cdots , n~.
\label{lambdaeqn}\end{equation}
It can be shown \cite{GPL} that the Lyapunov exponents are equal to
$\lambda_i/t$ in the limit $t \to \infty$. Thus, the Lyapunov
exponents can be obtained by solving the above differential equations
for long times.  However, since the right hand side depends on the
angles $\theta_i$, we also require differential equations governing
the evolution of these angles.

Differential equations for the angles can be obtained by comparing the
sub-diagonal elements in Eq. (\ref{finaleqn}). This gives
 $$
f_1({\dot{\theta}}) = S_{21};~f_2({\dot{\theta}}) = S_{31};~\cdots;
~f_{n(n-1)/2}({\dot{\theta}}) = S_{n,n-1}.  
$$ 
This set of differential equations can be transformed into a more
convenient form\cite{long}
\begin{equation}
\dot{\theta}_i = g_i({\theta}), \ \ \ i=1,2, \cdots ,n(n-1)/2~.
\label{thetaeqn}
\end{equation}
where the equations for $\theta_i$ are decoupled from the equations
for $\lambda_i$. This avoids potential problems with degenerate
Lyapunov spectra. Because of these reasons, the second method just
described is to be preferred over the method first
discussed. Equations (\ref{lambdaeqn}) and (\ref{thetaeqn}) form a
system of $n(n+1)/2$ ordinary differential equations that can be
solved to obtain the Lyapunov exponents.

Our system of differential equations has another attractive feature.
The equation for $\lambda_1$ depends only on the first $(n-1)$
$\theta_i$'s (under a suitable ordering) \cite{long}. Therefore, if
one is interested in only the largest Lyapunov exponent, one needs to
solve only $n$ equations (as opposed to $n(n+1)/2$ for the full
spectrum). The equation for $\lambda_2$ depends only on the first
$2n-3$ $\theta_i$'s. Therefore, to obtain the first 2 Lyapunov
exponents, one needs to solve only $2n-1$ equations. In general, to
solve for the first $m$ Lyapunov exponents, one has to solve
$m(2n-m+1)/2$ equations which is always less than $n(n+1)/2$ (the
total number of equations) for $m<n$. This is in contrast to the
situation for the conventional continuous QR or SVD methods where it
is computationally costlier to evaluate a partial spectrum once a
threshold is crossed \cite{GPL}. The first method discussed above
shares this disadvantage.

We end the general analysis of our system of equations by pointing out
an interesting fact. From Eq. (\ref{lambdaeqn}), 
\begin{equation}
\dot{\lambda}_1 + \dot{\lambda}_2 + \cdots + \dot{\lambda}_n = {\rm
Trace} (S)~. 
\end{equation}
Parametrizing the Jacobian matrix ${\bf DF}$ as $[{\bf DF}]_{ij} =
df_{ij}$ we can evaluate the trace of the matrix $S$ to obtain
\cite{long}
\begin{equation}
\dot{\lambda}_1 + \dot{\lambda}_2 + \cdots + \dot{\lambda}_n = 
df_{11} + df_{22} + \cdots + df_{nn}~.
\end{equation}
This relation can be used to speed up numerical integration of the
differential equation for $\lambda_n$. 

We now illustrate the second method for a simple two-dimensional
example. In this case, $Q$ is parametrized as follows: 
\begin{equation}
Q = \left(
\begin{array}{cc}
 \cos\theta_1 & \sin\theta_1 \\
             -\sin\theta_1 & \cos\theta_1
\end{array} \right)~,
\end{equation}
and the upper-triangular matrix $R$ may be written as,
\begin{equation}
R = \left(
\begin{array}{cc}
 e^{\lambda_1} & r_{12} \\
 0 & e^{\lambda_2}
\end{array} \right)~.
\end{equation}
Next, we parametrize the Jacobian matrix ${\bf DF}$ as follows:
\begin{equation}
{\bf DF} = \left( \begin{array}{cc}
df_{11} & df_{12} \\ df_{21} & df_{22}
\end{array} \right)~.
\end{equation}
Substituting the above into Eq. (\ref{finaleqn}), we obtain the
desired equations for $\lambda_1$, $\lambda_2$ and $\theta_1$: 
\begin{eqnarray}
\frac{d \lambda_1}{dt} & = & df_{11} \cos^2 \theta_1 + df_{22} \sin^2
\theta_1  \nonumber \\
 & & - \frac{1}{2} (df_{12} + df_{21}) \sin 2 \theta_1~, \nonumber \\
\frac{d \lambda_2}{dt} & = & df_{11} \sin^2 \theta_1 + df_{22}
\cos^2\theta_1 
\nonumber \\ 
 & & + \frac{1}{2} (df_{12} + df_{21}) \sin 2 \theta_1~, \label{eqs}\\
\frac{d \theta_1}{dt} & = & -\frac{1}{2}(df_{11}-df_{22}) \sin 2 \theta_1
\nonumber \\
 & &  + df_{12} \sin^2 \theta_1 - df_{21} \cos^2 \theta_1~. \nonumber
\end{eqnarray}
The above differential equations are numerically integrated forward in
time until the desired convergence for the exponents, $\lambda_1/t$
and $\lambda_2/t$, is achieved.

As our first example, we consider the driven van der Pol oscillator:
\begin{eqnarray}
\dot{z}_1 & = & z_2~,\\ \nonumber
\dot{z}_2& = &  -d (1-z_1^2) z_2 - z_1 + b \cos\omega t~.
\end{eqnarray}
For the parameter values $d = -5$, $b=5$, and $\omega=2.466$, our
results are in agreement with values obtained earlier using the
symplectic approach \cite{HR}.

To illustrate the application of the method to a system with more
degrees of freedom, we turn to the standard test case of the Lorenz
equations \cite{Lorenz}:
\begin{eqnarray}
\dot{z}_1&=&\sigma (z_2-z_1)~,\\ \nonumber
\dot{z}_2&=&z_1(\rho-z_3)-z_2~,\\ \nonumber
\dot{z}_3&=&z_1z_2-\beta z_3~.  \label{lorenzeqns}
\end{eqnarray}
For this three degree of freedom system, we need to generalize the
equations given in Eq. (\ref{eqs}). This can be easily done to obtain
six partially coupled differential equations governing the evolution
of the three Lyapunov exponents and three angles. We used parameter
values of $\sigma = 10$, $\rho = 28$, and $\beta = 8/3$. An extensive
comparison of our method against the standard QR method with
Gramm-Schmidt reorthogonalization (QR/GS) \cite{WEA} was carried
out. Both methods were applied to the same fiducial trajectory
generated using a RK4 integrator applied to Eqs. (\ref{lorenzeqns})
with time step $\epsilon=0.001$. Error and convergence analysis was
carried out by applying the two methods to the fiducial trajectory
sampled over time intervals $t_s\geq\epsilon$. Both methods were
implemented using RK4 integrators, and with $t_s=\epsilon=0.001$, both
generated essentially identical results. As a function of $t_s$, both
methods were quartically convergent as expected, QR/GS possessing a
smaller coefficient for the positive Lyapunov exponent and a larger
one for the negative exponent. Even for this small system, execution
times for both methods were similar. (We did not attempt to fully
optimize either one of the codes.) For larger systems our method is
expected to be more efficient.

In the Lorenz system of equations, the sum of the three Lyapunov
exponents must equal $-(\sigma + \beta +1)$. With our method, the sum
of the three Lyapunov exponents $-(\sigma + \beta +1) =
-13.6666\cdots$ was maintained to nine decimal places, {\em
independent} of the sampling interval over the investigated range,
$0.001\leq t_s\leq 0.02$, a property not shared by QR/GS. (The sum of
all Lyapunov exponents is an important quantity in stationary, thermostatted 
nonequilibrium systems since it is directly proportional to the transport
coefficients. Recent analytic and numerical results are reported in Refs. 
\cite{trans2}.)

To summarize, we have described a technique for computing Lyapunov
exponents that has several advantages over existing methods. The
minimal number of variables is used, rescaling and reorthogonalization
are eliminated, partial Lyapunov spectra can be calculated using a
fewer number of equations, there are no difficulties with degenerate
Lyapunov spectra, and global invariances of the Lyapunov spectrum can
be explicitly preserved. The method allows a natural fast/slow split
between variables which may be taken advantage of to improve
convergence of the exponents. Moreover, the simple form of the final
set of equations may prove to be useful in analytic
considerations as well.  Further details will be presented elsewhere
\cite{long}.

GR thanks Los Alamos National Laboratory for hospitality. His work was
supported in part by research grant from the Department of Science and
Technology, India. SH and RDR acknowledge support from the DOE Grand
Challenge award for Computational Accelerator Physics. Simulations
were performed on the SGI/Cray Origin 2000 at the ACL, LANL and the
Cray T3E at NERSC, LBNL. We thank the referees for valuable comments
which helped to greatly improve the paper.

\end{document}